\documentclass[12pt]{iopart}
\usepackage{amsfonts}
\usepackage{amssymb}
\usepackage{graphicx}
\usepackage{epstopdf}

\begin{document}

\title{Jamming and percolation of $k^2$-mers on simple cubic lattices}

\author{P. M. Pasinetti}
\address{Departamento de F\'{\i}sica, Instituto de F\'{\i}sica Aplicada (INFAP), Universidad Nacional de San Luis - CONICET, Ej\'ercito de Los Andes 950, D5700HHW, San Luis, Argentina}

\author{P. M. Centres}
\address{Departamento de F\'{\i}sica, Instituto de F\'{\i}sica Aplicada (INFAP), Universidad Nacional de San Luis - CONICET, Ej\'ercito de Los Andes 950, D5700HHW, San Luis, Argentina}

\author{A.J. Ramirez-Pastor}
\address{Departamento de F\'{\i}sica, Instituto de F\'{\i}sica Aplicada (INFAP), Universidad Nacional de San Luis - CONICET, Ej\'ercito de Los Andes 950, D5700HHW, San Luis, Argentina}

\ead{pmp@unsl.edu.ar}

\begin{abstract}
Jamming and percolation of square objects of size $k \times k$ ($k^2$-mers) isotropically deposited on simple cubic lattices have been studied by numerical simulations complemented with finite-size scaling theory. The $k^2$-mers were irreversibly deposited into the lattice. Jamming coverage $\theta_{j,k}$ was determined for a wide range of $k$ ($2 \leq k \leq 200$). $\theta_{j,k}$ exhibits a decreasing behavior with increasing $k$, being $\theta_{j,k\rightarrow\infty}=0.4285(6)$ the limit value for large $k^2$-mer sizes. On the other hand, the obtained results shows that percolation threshold, $\theta_{c,k}$, has a strong dependence on $k$. It is a decreasing function in the range $2 \leq k \leq 18$ with a minimum around $k=18$ and, for $k \geq 18$, it increases smoothly towards a saturation value. Finally, a complete analysis of critical exponents and universality has been done, showing that the percolation phase transition involved in the system has the same universality class as the 3D random percolation, regardless of the size $k$ considered.
\end{abstract}

\submitto{JOURNAL OF STATISTICAL MECHANICS: THEORY AND EXPERIMENTS}

\maketitle

\section{Introduction} \label{Introduction}


Random sequential adsorption (RSA) is one of the simplest model used for studying of irreversible adsorption processes \cite{Evans}.
An object of a given shape is placed randomly, sequentially and irreversibly on a substrate, subject to the constraint that it does not overlap previously deposited objects. The final state generated
by irreversible adsorption is a disordered state (known as jamming state), in which no more objects can be deposited due to the absence of free space of appropriate size and shape. The corresponding limiting or jamming coverage, $\theta_{j} \equiv  \theta(t=\infty)$ is less than that corresponding to the close packing ($\theta_{j} < 1$). Note that $\theta(t)$ represents the fraction of surface covered at time $t$ by the deposited objects.

If the concentration of the deposited objects on the substrate exceeds a critical value, a cluster (a group of occupied sites in such a way that each site has at least one occupied nearest neighbor site) extends from one side of the system to the other. This particular value of concentration rate is named critical concentration or percolation threshold $\theta_c$, and determines a phase transition in the system. This transition is a geometrical phase transition where the critical concentration separates a phase of finite clusters from a phase where an infinite cluster is present.

The percolation theory deals with the probability of occurrence of an infinite connectivity among the elements occupying on the lattice \cite{Stauffer}.
Thus, the jamming coverage has an important role on the percolation threshold, and the interplay between RSA and percolation is relevant for description of various deposition processes  \cite{Budi1,Budi2,Redner,Becklehimer,Vandewalle,Corne1,Corne2,Leroyer,Bonnier,Kondrat,Tara2012,Kondrat2017,EPJB4,PHYSA38}.
Objects with different shapes and sizes [e.g., linear \cite{Becklehimer,Vandewalle,Corne1,Corne2,Leroyer,Bonnier,Kondrat,Tara2012,Kondrat2017} and flexible \cite{Paw1,Paw2} $k$-mers (particles occupying $k$ adjacent sites), T-shaped objects and crosses \cite{Adam}, disks \cite{Connelly}, regular and star polygons \cite{Ciesla}, etc.] have been studied, and data of these studies show that the values of $\theta_j$ and $\theta_c$ strongly depend on the object shape and size.

In the case of square-shaped particles, which is the topic of this paper, the jamming and percolation problems have been studied in numerous works as useful objects for a description of both fundamental \cite{Nakamura86,Nakamura87,Centres2018,Feder,Brosilow,Privman,Rodgers,Mecke2002,Yamamoto2009,Shida,Carvalho,Kriuchevskyi} and practical problems \cite{Tahirkheli2007,Tahirkheli2010,Mitsen2016}.

In Refs. \cite{Nakamura86,Nakamura87,Centres2018}, the RSA problem of $k \times k$ square tiles ($k^2$-mers) on two-dimensional (2D) square lattices was studied by numerical simulations. The jamming coverage showed a decreasing behavior with increasing $k$, being $\theta_{j,k \rightarrow \infty}=0.5623(2)$ the limit value for large tile sizes. A finite-size scaling analysis of the jamming transition was carried out \cite{Centres2018}, and the corresponding spatial correlation length critical exponent $\nu_j$ was measured, being $\nu_j \approx 1$. In the same work, the obtained results for the percolation threshold revealed that $\theta_c$ is an increasing function of $k$ in the range $1 \leq k \leq 3$. For $k \geq 4$, all jammed configurations are non-percolating states, and consequently, the percolation phase transition disappears. This finding was corroborated by theoretical analysis based on exact calculations of all the possible configurations on finite cells. In addition, a complete analysis of critical exponents and universality have been done in Ref. \cite{Centres2018}, showing that the percolation phase transition involved in the system has the same universality class as the ordinary random percolation, regardless of the size $k$ considered.


In contrast to the statistic for the simple particles, the degeneracy of arrangements of extended objects is strongly influenced by the structure and dimensionality of the lattice. In this context, the present paper deals with jamming and percolation aspects of $k \times k$ square plaquettes deposited on 3D simple cubic lattices. Using extensive simulations supplemented by finite-size scaling analysis, jamming coverage and percolation thresholds were determined for a wide range of $k$ values. The obtained results allow us to report the functionality of jamming coverage and percolation threshold with the object size. In addition, the accurate determination of the critical exponents indicate that the percolation transition of $k^{2}$-mers on simple cubic lattices belongs to the 3D random percolation universality, and that the jamming transition can be characterized by an exponent $\nu_{j} = 2/3$.

The paper is organized as it follows: the model is presented in Section \ref{modelo}. Jamming and percolation properties are studied in Sections \ref{jam} and \ref{perco}, respectively. Finally, the conclusions are drawn in Section \ref{conclu}.

\section{The model}\label{modelo}

Let us consider the substrate represented by a 3D simple cubic lattice of $M=L \times L \times L$ sites (an $L^3$-lattice) with periodic boundary conditions in each direction (a torus). In this way, all the lattice sites are equivalent and there are no edge effects in the deposition process. The filling of the lattice with $k^{2}$-mers (objects ocupping $k \times k \times 1$ sites) is carried out following the conventional $RSA$ process \cite{Evans}. It consists of three steps, namely, (i) starting from an initially empty lattice; (ii) then, a square tile of $k \times k \times 1$ sites is chosen at a random position and orientation and, if those sites are empty, a $k^{2}$-mer is deposited on them; otherwise, the attempt is rejected; (iii) steps $(i)-(ii)$ are repeated until a desired concentration $\theta=k^{2}N/M$ is reached ($N$ is the number of the deposited $k^{2}$-mers).

\section{Jamming coverage}\label{jam}

As mentioned in Section \ref{Introduction}, due to the increasing probability of blocking on the lattice by the already randomly deposited objects, the jamming coverage is less than the close-packing one ($\theta_j < 1$). Consequently, $\theta$ ranges from 0 to $\theta_j$ for objects occupying more than one site, and the interplay between jamming and percolation must be considered.

For the purpose of obtaining the jamming threshold as a function of $k$, the probability $W_{L,k}(\theta)$ that an $L^3$-lattice reaches a coverage $\theta$ has been calculated taking into account the numerical method introduced in Ref. \cite{EPJB4}. According to it, starting with an initially empty $L^3$-lattice, a deposition process of $k^{2}$-mers is carried out until a particular jamming state has  been reached. $n$ runs of such process were carried out for each lattice size $L$. Then, the probability was calculated as: $W_{L,k}(\theta)=n_L(\theta)/n$, where $n_L(\theta)$ is the number of samples that reach a coverage $\theta$. A set of $n=10^5$ independent samples were numerically prepared for several values of $L/k$=4, 6, 8, 10 and 20. The ratio $L/k$ was kept constant to avoid spurious results.


\begin{figure}
  \begin{center}
  \includegraphics[scale=0.4]{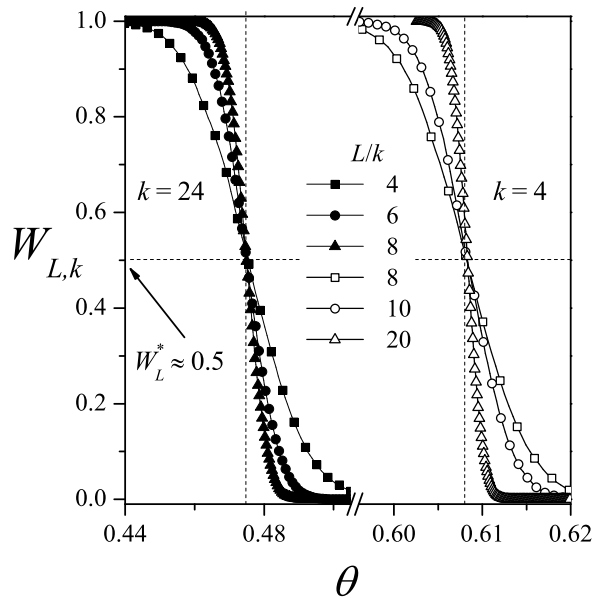}
  \end{center}
\caption{ Curves of the jamming probability $W_{L,k}$ as a function of the fraction of occupied sites $\theta$ for two values of tile size, $k$=4 and $k$=24. For clarity, three sizes are shown in the figure for each $k$, as indicated.}
\label{fig1}
\end{figure}

In Fig. \ref{fig1}, the curves of probability for the different $L/k$ values are shown for two typical cases, $k=4$  and 24. As mentioned in the previous paragraph, the simulations were performed for lattice sizes ranging between $L/k = 4$ and $L/k = 20$. For clarity, three sizes are shown in the figure for each $k$. With independence of the size $k$, the curves $W_{L,k}(\theta)$ approach to the step function as $L$ grows to infinity. Alternatively, for a finite value of $L$, the probability $W_{L,k}(\theta)$ varies continuously from 1 to 0. From the inspection of Fig. \ref{fig1}, it can be seen that: (i) for each tile size $k$, the curves cross each other in a nontrivial value $W^{*}_{k}$; (ii) those points are located at very well defined values in the $\theta$-axes determining the jamming threshold for each $k$ ($\theta_{j,k}$), (iii) $\theta_{j,k}$ decreases for increasing values of $k$.

\begin{figure}
\begin{center}
  \includegraphics[scale=0.4]{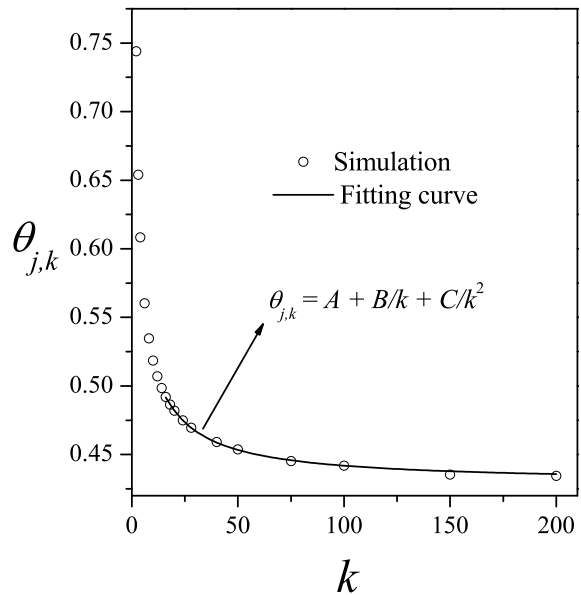}
  \end{center}
\caption{Jamming coverage $\theta_{j,k}$ as a function of $k$ for $k^{2}$-mers on 3D simple cubic lattices with $k$ between 2 and 200 (open circles). The solid
line corresponds to the fitting function as discussed in the text.}
\label{fig2}
\end{figure}

The procedure of Fig. \ref{fig1} was repeated for $k$ from 2 to 200, the results are presented in Fig. \ref{fig2} and collected in Table 1. From $k \geq $12 the data have been fitted by the function $\theta_{j,k}$=$A+B/k+C/k^{2}$, as proposed in Ref.\cite{Bonnier}; it is found that  $A=\theta_{j,k\rightarrow\infty}$=0.4285(6), $B$=1.30(4) and $C$=-4.4(4). To the best of our knowledge the value $\theta_{j,k=\infty}$=0.4285(6) has not been reported up to now.

The decreasing behavior of the jamming coverage with the size $k$ towards an asymptotic limit value has been already observed in numerous systems. The cases of linear $k$-mers \cite{Bonnier} or tiles \cite{Nakamura86} on square lattices, linear $k$-mers on triangular lattices \cite{Ernesto}, or $k$-mers on the 3D simple cubic lattice \cite{EPJB4}, are examples of this.

The value $\theta_{j,k\rightarrow\infty}$=0.5623(2) for $k^2$-mers on the 2D square lattice \cite{Nakamura86} is less than the value $\theta_{j,k\rightarrow\infty}$=0.660(2) obtained for linear $k$-mers in the same geometry \cite{Bonnier}. It means that less compact objects like linear $k$-mers are more effective in filling the square lattice than $k \times k$ square tiles. In 3D systems the same trend seems to be maintained, at least for small object sizes. As an illustrative example, in the simple cubic lattice we have $\theta_{j,k=20}$=0.5256 for linear $k$-mers \cite{EPJB4} and $\theta_{j,k=20}$=0.4820 for $k^2$-mers (tiles). However, this seem not to be valid for large values of $k$. Thus, $\theta_{j,k\rightarrow\infty}$=0.4045(19) for $k$-mers whereas $\theta_{j,k\rightarrow\infty}$=0.4285(6) for $k^2$-mers. The limiting values of $\theta_{j,k}$ were obtained by simulations for relatively small $k$ sizes and then extrapolated to represent very long objects. Additional simulation research of RSA with extremely long objects should be performed in the future to confirm or reject the prediction in this point.

In order to complete the jamming study, the critical exponent $\nu_j$ of the jamming transition was obtained. For this purpose, it is useful to define the quantity $W_{L,k}^{'}=1-W_{L,k}$, which is fitted by the error function because $dW^{'}_{L,k}/d \theta$ is expected to behave like the Gaussian distribution \cite{Vandewalle},
\begin{equation}\label{jamming }
    \frac{dW_{L,k}^{'}}{d\theta}=\frac{1}{\sqrt{2\pi}\Delta_{L,k}^{'}}\exp \left\{ -\frac{1}{2} \left[\frac{\theta-\theta_{j,k}(L)}{\Delta_{L,k}^{'}}
    \right]^2 \right\},
\end{equation}
where $\theta_{j,k}(L)$ is the concentration at which the slope of $dW_{L,k}^{'}/d \theta$ is the largest
and $\Delta_{L,k}^{'}$ is the standard deviation from $\theta_{j,k}(L)$.

Then, $\nu_j$ can be calculated from the maximum of $dW_{L,k}^{'}/d\theta$:
\begin{equation}\label{derimax_jam}
  \left(\frac{dW_{L.k}^{'}}{d\theta}\right)_{\rm max}\propto L^{1/\nu_j}.
\end{equation}

\begin{figure}
\begin{center}
  \includegraphics[scale=0.4]{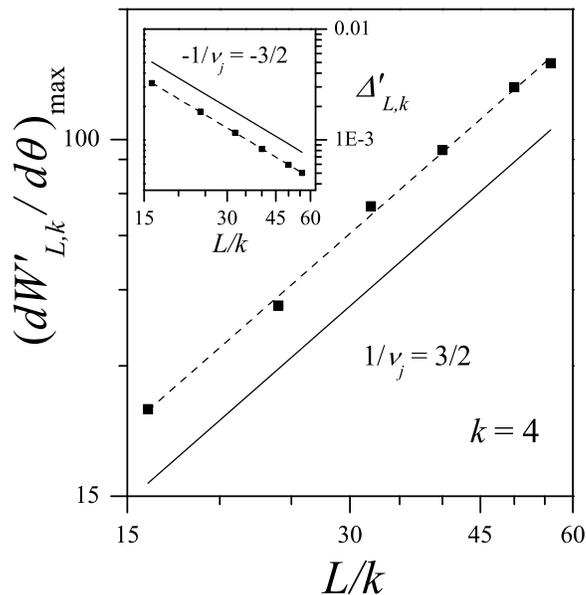}
  \end{center}
\caption{Log-log plot of $(dW_{L,k}^{'}/d\theta)_{\rm max}$ as a function of $L/k$ for $k$=4. According to Eq. (\ref{derimax_jam}) the slope of the line corresponds to 1/$\nu_{j}$. Inset: Log-log plot of the standard deviation $\Delta_{L,k}^{'}$ in Eq. (\ref{Deltamax_jam}) as a function of $L/k$ for the same case shown in part (a). According to Eq. (\ref{Deltamax_jam}), the slope of the line corresponds to -1/$\nu_{j}$.}
\label{fig3}
\end{figure}

Figure \ref{fig3} shows, in a log-log scale, $({dW_{L,k}^{'}}/{d\theta})_{\rm max}$ as a function of $L/k$ for $k$=4, where $\nu_j$ can be obtained from the inverse of the slope of the line that fits the data, in this case $\nu_j$=0.67(2).

An alternative way to obtain $\nu_j$ is from the divergence of the root mean square deviation of the jamming observed from their average values,
$\Delta_{L,k}$,
\begin{equation}\label{Deltamax_jam}
  \Delta_{L,k}^{'}\propto L^{-1/\nu_j}.
\end{equation}

In this case, the slope of the fitting line for $\Delta_{L,k}^{'}$ versus $L/k$ in log-log scale corresponds to -1$/\nu_j$. The inset in Fig. \ref{fig3} shows $\log (\Delta_{L,k}^{'})$ as a function of $\log(L/k)$ for the same case of the main figure.  Again, the obtained value for the critical exponent, $\nu_j$=0.67(1), remains close to 2/3.

The procedure in Fig. \ref{fig3} (and the corresponding inset) was repeated for different values of $k$. In all cases, the value obtained for $\nu_j$ remains close to 2/3. To the best of our knowledge, this value has not been reported up to now.

\begin{table}
\begin{center}
$$
\begin{array}{|c|c|}
 \hline
k	&	\theta_{j,k}	\\
\hline
2	&	0.7439(9)	\\
\hline
3	&	0.6540(8)	\\
\hline
4	&	0.6083(2)	\\
\hline
6	&	0.5603(6)	\\
\hline
8	&	0.5347(6)	\\
\hline
10	&	0.5185(6)	\\
\hline
12	&	0.5070(17)	\\
\hline
14	&	0.4985(17)	\\
\hline
16	&	0.4919(17)	\\
\hline
18	&	0.4864(16)	\\
\hline
20	&	0.4820(16)	\\
\hline
24	&	0.4749(17)	\\
\hline
28	&	0.4696(23)	\\
\hline
40	&	0.4591(24)	\\
\hline
50	&	0.4537(24)	\\
\hline
75	&	0.4453(38)	\\
\hline
100	&	0.4419(74)	\\
\hline
150	&	0.435(12)	\\
\hline
200	&	0.435(11)	\\
\hline
\end{array}
$$
\caption{Numerical values of jamming coverage $\theta_{j,k}$ as a function of $k$. Error estimates concerning the last digits are indicated between parentheses.}
\end{center}
\end{table}

\section{Percolation}\label{perco}

\subsection{Calculation method and percolation thresholds}

According to the percolation theory, the central idea rests on finding the minimum concentration $\theta$ = $\theta_c$ for which at least one cluster emerges connecting the opposite sides of the system. In our case we will study: i) the percolation threshold as a function of the size of the tiles $\theta_{c,k}$, and ii) the universality class of the phase transition.

To achieve the two points above mentioned, the basic procedure provided by finite-size scaling theory was used. For this reason, different probabilities of percolation were calculated as well as the percolation order parameter and its corresponding susceptibility for different system sizes.

Let $R=R^{X}_{L,k}(\theta)$ represents the probability that a lattice $L \times L \times L$  percolates at the concentration $\theta$ by the deposition of $k \times k \times 1$ tiles \cite{Yone1}. According to our analysis, $X$ may have the following meanings:

\begin{itemize}
  \item $R^{R}_{L,k}(\theta)$: the probability of finding a rightward percolating cluster, along the $x$-direction,\\
  \item $R^{D}_{L,k}(\theta)$: the probability of finding a downward percolating cluster, along the $z$-direction,\\
  \item $R^{F}_{L,k}(\theta)$: the probability of finding a frontward percolating cluster, along the $y$-direction.
\end{itemize}

Other useful definitions for the finite-size analysis are:

\begin{itemize}
  \item $R^{U}_{L,k}(\theta)$: the probability of finding a cluster which percolates on any direction,\\
  \item $R^{I}_{L,k}(\theta)$: the probability of finding a cluster which percolates in the three (mutually perpendicular) directions,\\
  \item $R^{A}_{L,k}(\theta)$=$\frac{1}{3}[R^{R}_{L,k}(\theta)+R^{D}_{L,k}(\theta)+R^{F}_{L,k}(\theta)]$: the arithmetic average.
\end{itemize}

Through computational simulation, each of the previously mentioned quantities were calculated. Basically, each simulation consists of the following steps: $(a)$ the construction of a simple cubic lattice of linear size $L$ with a coverage $\theta$, $(b)$ the cluster analysis using the Hoshen and Kopelman algorithm \cite{Hoshen} with open boundary conditions, and $(c)$ the determination of the largest cluster size $S_L$.

A total of $m_L$ independent runs of such two steps procedure were carried out for each lattice size $L$. Then, the probabilities has been calculated as: $R^{X}_{L,k}(\theta)=m^{X}_{L}/m_L$, where $m^{X}_L$ indicates the number of percolating samples.

The percolation order parameter and the corresponding susceptibility $\chi$ and reduced fourth-order cumulant $U_L$ have been obtained from the largest cluster size \cite{Binder,Biswas,Chandra}. Thus,
\begin{equation}
    P=\frac{<S_L>}{M},
\end{equation}
\begin{equation}
    \chi=\frac{\left[  \langle S^{2}_{L}\rangle -  \langle S_{L}\rangle^{2} \right]}{M},
\end{equation}
and
\begin{equation}\label{cum}
U_L=1-\frac{\langle S_{L} ^4\rangle}{3\langle
S_{L}^2\rangle ^2},
\end{equation}
where  $<..>$ means an average over simulation runs.

\begin{figure}
\begin{center}
\includegraphics[scale=0.4]{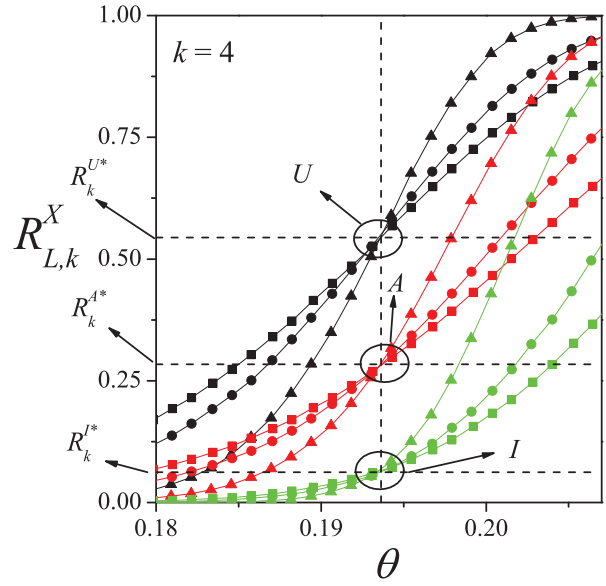}
\end{center}
\caption{Fraction of percolating lattices $R^{X}_{L,k}(\theta)$ ($X=\{I,U,A\}$, as indicated) as a function of the concentration $\theta$ for $k=4$ and different lattice sizes: $L/k$=12, squares; $L/k$=15, circles; $L/k$=24 triangles up. The statistical error is smaller than the symbol size. \label{fig4}}
\end{figure}

In the percolation simulations, a total of $m_L=10^{5}$ independent samples have been used to calculate averages. In addition, for each value of $k$, the finite size scaling study was carried out by using the values $L/k = 4, 6, 8, 10,12, 15$ and $24$. As it can be appreciated, this represents extensive calculations from the computational point of view. From this analysis, the percolation threshold and the critical exponents can be determined with reasonable accuracy.

\begin{figure}
\begin{center}
\includegraphics[scale=0.4]{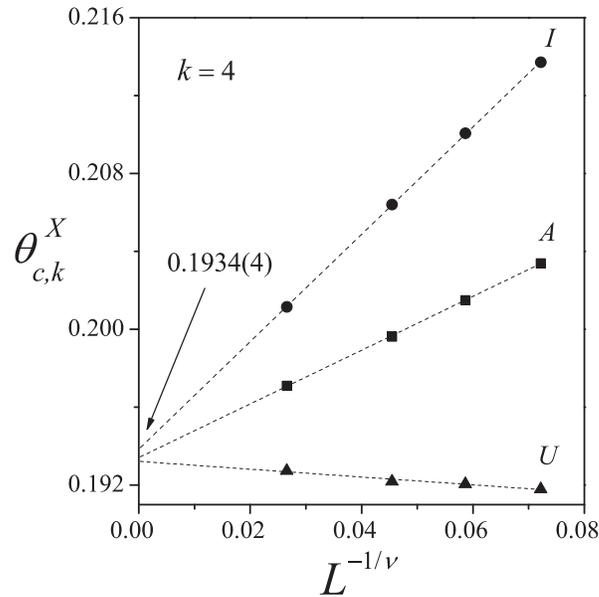}
\end{center}
\caption{Extrapolation of the percolation threshold for an $L^3$-lattice $\theta^{X}_{c,k^{}}(L)$ $(X=\{I,U,A\})$ towards the thermodynamic limit according to the theoretical prediction given by Eq. (\ref{limit}). Circles, squares and triangles denote the values of $\theta^{X}_{c,k^{}}(L)$ obtained by using the criteria $I$, $A$ and $U$, respectively, for $k=$4.}
\label{fig5}
\end{figure}

\begin{table}
\begin{center}
$$
\begin{array}{|c|c|}
 \hline
k	&	\theta_{c,k}	\\
\hline
2	&	0.24077	0   \\
\hline
3	&	0.2103(5)	\\
\hline
4	&	0.1934(4)	\\
\hline
5	&	0.1819(5)	\\
\hline
6	&	0.1742(7)	\\
\hline
7	&	0.1682(13)	\\
\hline
8	&	0.1641(8)	\\
\hline
9	&	0.1606(11)	\\
\hline
10	&	0.1578(6)	\\
\hline
11	&	0.1557(14)	\\
\hline
12	&	0.1548(9)	\\
\hline
14	&	0.1534(16)	\\
\hline
20	&	0.1520(17)	\\
\hline
24	&	0.1527(19)	\\
\hline
32	&	0.1559(22)	\\
\hline
48	&	0.1618(19)	\\
\hline
64	&	0.1656(29)	\\
\hline
80	&	0.1675(38)	\\
\hline
100	&	0.170(5)	\\
\hline
150	&	0.175(4)    \\
\hline
200	&	0.177(5)	\\
\hline
\end{array}
$$
\caption{Numerical values of percolation threshold $\theta_{c,k}$ as a function of $k$. Error estimates concerning the last digits are indicated between parentheses.}
\end{center}
\end{table}

The theory of finite-size scaling \cite{Stauffer,Yone1,Binder} gives us an efficient way to estimate the percolation threshold from the maximum of the curves of $R^{X}_{L,k}(\theta)$ (see Fig. \ref{fig4}). For this, the different curves are expressed as a function of continuous values of $\theta$. Then, as in the case of the jamming probability, $dR^{X}_{L,k}/d \theta$ can be approximated by the Gaussian function. We use the term approximated because the behavior of $dR^{X}_{L,k}/d \theta$ is known not to be a Gaussian in all range of coverage \cite{Newman}. However, this quantity is approximately Gaussian near the peak, and fitting with a Gaussian function is a good approximation for the purpose of locating its maximum. Thus,
\begin{equation}\label{ecu1}
    \frac{dR^{X}_{L,k}}{d\theta}=\frac{1}{\sqrt{2\pi}\Delta^{X}_{L,k}}\exp \left\{ -\frac{1}{2} \left[\frac{\theta-\theta_{c,k}^{X}(L)}{\Delta^{X}_{L,k}}
    \right]^2 \right\},
\end{equation}
where $\theta^{X}_{c,k}(L)$ is the concentration at which the slope of $dR^{X}_{L,k}/d \theta$ is the largest
and $\Delta^{X}_{L,k}$ is the standard deviation from $\theta^{X}_{L,k}(L)$.

Once the values of $\theta^{X}_{c,k}(L)$ were obtained for all lattice sizes, the percolation thresholds were calculated by scaling analysis \cite{Stauffer}. In this way, the following relationship is got
\begin{equation}\label{limit}
    \theta^{X}_{c,k}(L)=\theta^{X}_{c,k}(\infty)-A^{X}L^{-1/\nu},
\end{equation}
where $A^{X}$ is a nonuniversal constant and $\nu$ is the critical exponent of the
correlation length which has been taken as $7/8$ for the present, since, as it will
be shown below, our model belongs to the same universality class as random 3D percolation \cite{Stauffer}.

Figure \ref{fig5} shows the extrapolation toward the thermodynamic limit of $\theta^{X}_{c,k}(L)$ ($X = {I,U,A}$ and $k=4$) according to Eq. (\ref{limit}). Then, the final values of $\theta^{X}_{c,k}(\infty)$ are given as: $\theta_{c,k} \pm \delta_k$, where $\delta_k= {\rm max}(\mid \theta^{U}_{c,k} - \theta^{A}_{c,k}\mid, \mid \theta^{I}_{c,k} - \theta^{A}_{c,k}\mid)$. The values obtained in Fig. \ref{fig5} were: $\theta_{c,k=4}(\infty)=0.1934(4)$. For the rest of the paper, we will denote the percolation threshold for each size $k$ by $\theta_{c,k}$ [for simplicity we will drop the symbol``$(\infty)$"].

\begin{figure}
\begin{center}
\includegraphics[scale=0.4]{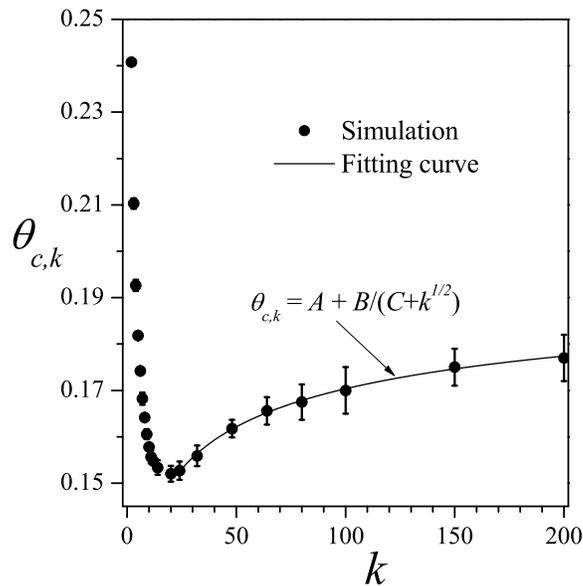}
\end{center}
\caption{The percolation threshold $\theta_{c,k}$ as a function of $k$ for $k^{2}$-mers on 3D simple cubic lattices. The minimum is reached around $k=18$. Symbols represent simulation results and solid line corresponds to the fitting function as discussed in the text.}
\label{fig6}
\end{figure}

The procedure of Fig. \ref{fig5} was repeated for $k$ ranging from 2 to 200. The results are shown in  Fig. \ref{fig6} and collected in Table 2. As can be seen from the figure, $\theta_{c,k}$ shows a nonmonotonic dependence with $k$, decreasing for small sizes, going through a minimum around $k \approx 18$, and finally slowly increasing for $k \geq 18$. 

For $k \rightarrow \infty$, the $\theta_{c,k}$ curve seems to tend toward a saturation value. In order to calculate this limit value, the simulation data were fitted with the function proposed in Ref. \cite{Slutskii},
\begin{equation}\label{extra}
 \theta_{c,k}= A + \frac{B}{C+\sqrt{k}} \ \ \ \ \ \ \ \ \ (k \geq 24) ,
\end{equation}
where $A=0.201(4)$, $B=-0.42(8)$, $C=3.82(1.07)$ and the adjusted coefficient of determination is $R^2=0.998$. Thus, $A=\theta_{c,k \rightarrow \infty}=0.201(4)$ represents the percolation threshold for infinitely large $k^2$-mers on simple cubic lattices. This limit value has been derived by using an extrapolation method [Eq. (\ref{extra})], and more extensive simulations are necessary for obtaining a direct confirmation of the percolation behavior of extremely large tiles.

\subsection{Critical exponents and universality}

In this section, the critical exponents $\nu$, $\beta$ and $\gamma$ will be calculated.
Knowing $\nu$, $\beta$ and $\gamma$ is enough to determine the universality class of
our system and understand the related phenomena.

The standard theory of finite size \cite{Binder} allows us for various routes to estimate
the critical exponent $\nu$  from simulation data. One of these methods is from the maximum
of the function ${dR^{X}_{L,k}}/{d\theta}$,
\begin{equation}\label{derimax}
  \left(\frac{dR^{X}_{L,k}}{d\theta}\right)_{max} \propto L^{1/\nu}.
\end{equation}

\begin{figure}
\begin{center}
\includegraphics[scale=0.5]{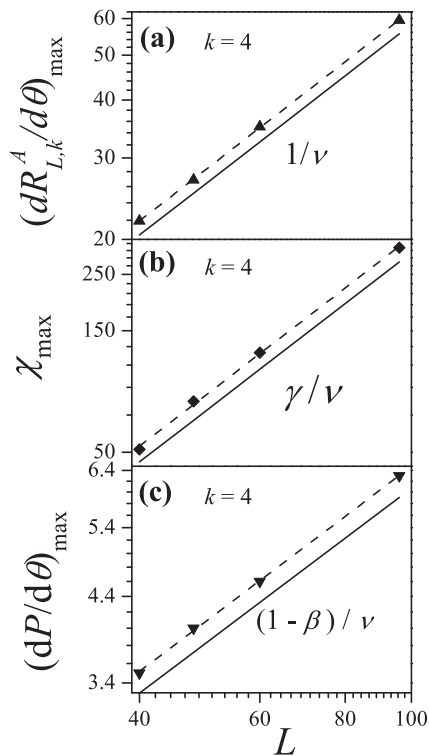}
\end{center}
\caption{a) Maximum of the derivative of the $A$ percolation probability, $\left(d R^{A}_{L,k}/d \theta \right)_{\rm max}$ as a function of $L/k$ (in a log-log scale) for $k=4$. b) Idem for susceptibility. c) Idem for the maximum of the order parameter derivative.}
\label{fig7}
\end{figure}

In Fig. \ref{fig7}(a), $\log \left[ \left(dR^{X}_{L,k}/d\theta\right)_{\rm max}\right]$ has been plotted as a function of $\log [L]$ for $k$=4. According to Eq. (\ref{derimax}), the slope of the fitting line corresponds to $1/\nu$. As can be observed in all the cases, the data present a fairly linear behaviour, giving the value $\nu=0.875(5)$.

\begin{figure}
\begin{center}
\includegraphics[scale=0.3]{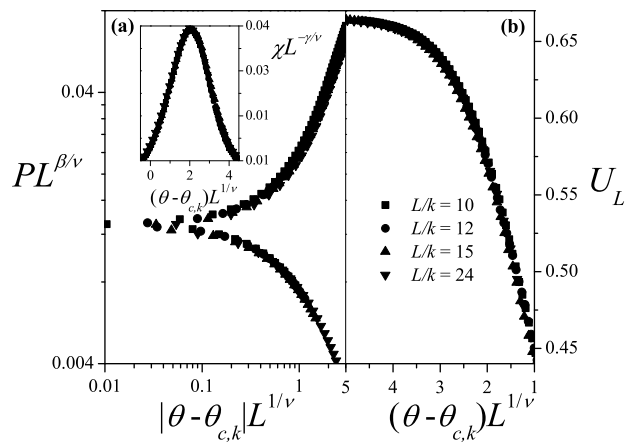}
\end{center}
\caption{a) Data collapse for the order parameter, corresponding to $k$=4. Inset: idem for the susceptibility. b) Idem for the Binder cumulant.}
\label{fig8}
\end{figure}

Obtaining $\nu$ make us able to estimate the exponents $\gamma$ and $\beta$ according to the theory \cite{Stauffer}. In the case of $\gamma$, it is obtained by scaling the maximum value of the susceptibility, according to the scaling assumption for this quantity given by $\chi=L^{\gamma/\nu}\overline{\chi}(u)$, where $u=(\theta-\theta_{c,k})L^{1/\nu}$ and $\overline{\chi}$ is the corresponding scaling function. At the point where $\chi$ is maximal, $u$=const. and $\chi_{\rm max} \propto L^{\gamma/\nu}$. The simulation data are shown in Fig. \ref{fig7}(b). From a linear fit, the obtained value for the exponent is $\gamma=1.81(3)$.

On the other hand, the exponent $\beta$ is calculated from the scaling behavior of the order parameter at criticality, $P=L^{-\beta/\nu}\overline{P}(u')$, where $u'=|\theta-\theta_{c,k}|L^{1/\nu}$ and $\overline{P}$ is the scaling function. At the point where $dP/d\theta$ is maximal, $u'$=const. and,

\begin{equation}\label{Deripara}
    \left(\frac{dP}{d\theta}\right)_{\rm max}=L^{(-\beta/\nu+1/\nu)}{\overline{P}(u')\propto L^{(1-\beta)/\nu}}.
\end{equation}

The scaling of $(dP/d\theta)_{\rm max}$ is shown in Fig. \ref{fig7}(c).
From the slope of the fitting line, the obtained value for the exponent is $\beta=0.42(2)$.

The procedure showed in Fig. \ref{fig7} was repeated for different sizes $k$ ranging between 2 and 200. In all the cases, the values for the exponents $\nu$, $\beta$ and $\gamma$
agree very well with the known values for 3D random percolation: $\nu \approx 0.8774$ \cite{Koza},  $\beta \approx 0.4273 $ \cite{Gracey} and $\gamma \approx 1.8357$ \cite{Gracey}. See Wikipedia webpage: https://en.wikipedia.org/wiki/Percolation$_-$critical$_-$exponents.

Finally, the scaling behavior can be further tested by plotting $PL^{\beta/\nu}$ versus $|\theta-\theta_{c,k}|L^{1/\nu}$
,$\chi L^{-\gamma/\nu}$ versus $(\theta-\theta_{c,k})L^{1/\nu}$ and $U$ versus $(\theta-\theta_{c,k})L^{1/\nu}$
and looking for data collapsing. The results are showed in Figs. \ref{fig8}(a) and \ref{fig8}(b), using the values of $\theta_{c,k}$ obtained and the values of the critical exponents corresponding to ordinary 3D percolation.
As can be seen, the data scaled extremely well, supporting the hypothesis that the model belongs to the universality class of the 3D random percolation.

\section{Conclusions}  \label{conclu}

The behavior of jamming and percolation thresholds in RSA of $k \times k$ square objects ($k^2$-mers) deposited on simple cubic lattices have been studied by numerical simulations complemented with finite-size scaling theory.

The dependence of the jamming coverage $\theta_{j,k}$ on the size $k$ was studied for $k$ ranging from 2 to 200. A decreasing behavior was observed for $\theta_{j,k}$, with a finite value of saturation in the limit of infinitely long $k^2$-mers: $\theta_{j,k}= A + B/k + C/k^2$ $(k \geq 12)$, being $A=\theta_{j,k\rightarrow\infty}=0.4285(6)$, $B$=1.30(4) and $C$=-4.4(4). The value $\theta_{j,k=\infty}=0.4285(6)$ is reported for the first time in the literature.

A decreasing behavior was also found for RSA of linear $k$-mers \cite{PHYSA38} on simple cubic lattices. However, some important differences between these systems can be observed: $(1)$ in the range of small sizes ($2 \leq k \leq 50$), the linear $k$-mers are more effective in filling the 3D cubic lattice than the $k \times k$ tiles; and $(2)$ the tendency described in point $(1)$ seems to become invalid for large values of $k$, being $\theta_{j,k=\infty}$=0.4045(19) \cite{PHYSA38} and  0.4285(6), for linear $k$-mers and $k^2$-mers, respectively. 

Based on scaling properties of the jamming probability $W_{L,k}(\theta)$, the critical exponent $\nu_j$ was measured for different object sizes $k$. In all cases, the values obtained for $\nu_j$ remain close to 3/2. This value differs clearly from the value $\nu_j \approx 1$ reported by Vandewalle et al. \cite{Vandewalle} for the case of linear $k$-mers on square lattices, and from other 2D systems \cite{Centres2018,JSTAT9}.

A nonmonotonic size dependence was found for the percolation threshold $\theta_{c,k}$, which decreases for small particles sizes, goes through a minimum around $k \approx 18$, and finally asymptotically converges towards a definite value for large sizes $k$. The simulation data were fitted with the function proposed in Ref. \cite{Slutskii}, $\theta_{c,k}= A + B/(C+\sqrt{k})$ $(k \geq 24)$, where $A=\theta_{c,k \rightarrow \infty}=0.201(4)$ represents the percolation threshold for infinitely large $k^2$-mers on simple cubic lattices. A similar behavior was reported recently in the case of linear $k$-mers on 2D square lattices \cite{Slutskii}. A common feature in these systems is that in both cases $(d-1)$-dimensional objects are deposited on $d$-dimensional substrates. Future efforts will be made to study other systems of linear and planar objects on 2D and 3D lattices. This will allow us to explore and discuss the obtained percolation properties in terms of the relationship between the dimension of the depositing object and the dimension of the
substrate.

Finally, the accurate determination of critical exponents ($\nu$, $\gamma$ and $\beta$) revealed that the model belongs to the same universality class as the 3D random percolation, regardless of the size $k$ considered.

\section{ACKNOWLEDGMENTS}

This work was supported in part by CONICET (Argentina) under project number PIP 112-201101-00615; Universidad Nacional de San Luis (Argentina) under project No. 03-0816; and the National Agency of Scientific and Technological Promotion (Argentina) under project  PICT-2013-1678. The numerical work were done using the BACO parallel cluster (http://cluster\_infap.unsl.edu.ar/wordpress/) located  at Instituto de F\'{\i}sica Aplicada, Universidad Nacional de San Luis - CONICET, San Luis, Argentina.

\end{document}